\documentstyle[graphicx,amsmath,amssymb,fleqn]{mn}

\newcommand{\txd}{\text{d}}

\newcommand{\bfr}{{\boldsymbol{r}}}
\newcommand{\bfn}{{\boldsymbol{n}}}

\renewcommand{\leq}{\leqslant}
\renewcommand{\geq}{\geqslant}

\title[Radiative transfer in disc galaxies I]%
{Radiative transfer in disc galaxies I -- A comparison of \\ four methods
to solve the transfer equation in plane-parallel geometry}

\author[Baes \& Dejonghe]{Maarten Baes\thanks{Research Assistent of
the Fund of Scientific Research -- Flanders (Belgium)} and Herwig
Dejonghe\\ Sterrenkundig Observatorium Universiteit Gent, Krijgslaan
281 S9, B-9000 Gent, Belgium, maarten.baes@rug.ac.be}

\begin{document}

\maketitle
\begin{abstract}
Accurate photometric and kinematic modelling of disc galaxies requires
the inclusion of radiative transfer models. Due to the complexity of
the radiative transfer equation (RTE), sophisticated techniques are
required. Various techniques have been employed for the attenuation in
disc galaxies, but a quantitative comparison of them is difficult,
because of the differing assumptions, approximations and accuracy
requirements which are adopted in the literature. In this paper, we
present an unbiased comparison of four methods to solve the RTE, in
terms of accuracy, efficiency and flexibility. We apply them all on
one problem that can serve as a first approximation of large portions
of disc galaxies: a one-dimensional plane-parallel geometry, with both
absorption and multiple scattering taken into account, with an
arbitrary vertical distributions of stars and dust and an arbitrary
angular redistribution of the scattering. We find that the spherical
harmonics method is by far the most efficient way to solve the RTE,
whereas both Monte Carlo simulations and the iteration method, which
are straightforward to extend to more complex geometries, have a cost
which is about 170 times larger.
\end{abstract}


\section{Introduction}

In order to study the structure of galaxies, it is necessary to obtain
their intrinsic three-dimensional light distribution, i.e.\ to
deproject their two-dimensional image. This deprojection is
complicated by the effects of interstellar dust, which change the
radiation field on its way through the galaxy by absorption, emission
and scattering. The amount of interstellar dust present in disc
galaxies has been the subject of debate for the last few years. The
most widely supported idea is that spiral galaxies are moderately
optically thick in their central regions (a face-on optical depth of
order unity in the $V$ band), and optically thin in the outer
regions. This idea is supported by statistical studies (Giovanelli et
al.\ 1994, Boselli \& Gavazzi~1994, Buat \& Burgarella~1998), studies
of the extinction of galaxies in overlapping pairs (White \& Keel
1992, Berlind et al.\ 1997, R\"onnback \& Shaver~1997, Pizagno \&
Rix~1998, White et al.\ 2000), and detailed modelling of the
extinction in individual galaxies (Jansen et al.\ 1994, Ohta \&
Kodaira 1995, Xilouris et al.\ 1997, 1998, 1999, Kuchinski et al.\
1998). However, there is still no consensus, and other authors claim
that spiral galaxies are optically thick all over their disc
(Valentijn 1990, 1994, Burnstein et al.\ 1991, James \& Puxley~1993,
Peletier et al.\ 1995). Moreover other arguments complicate the
discussion: it is a gross simplification to talk about {\em the}
opacity of spiral galaxies, because there can be a large difference in
opacity between arm and interarm regions (White et al.\ 2000). For a
detailed overview of the opacity of disc galaxies we refer to Davies
\& Burnstein~(1995) and Kuchinski et al.\ (1998).

Many authors have demonstrated that dust attenuation (i.e.\ the
combined effect of absorption and scattering) has considerable effects
on photometric properties such as magnitudes, colors and scalelengths
(Witt et al.\ 1992, Byun et al.\ 1994, Corradi et al.\ 1996). These
effects are of a complicated nature, and are not simply proportional
to the optical depth of the galaxy. Also kinematic studies can be
complicated by dust attenuation because the projected kinematics will
be biased towards the motions of stars on the near side of the
line-of-sight (Bosma et al.\ 1992, Matthews \& Wood~2000, Baes \&
Dejonghe~2001a). Both photometric and kinematic studies of disc
galaxies hence require sophisticated deprojection techniques which
take dust attenuation into account. The only means to do this properly
is by constructing radiative transfer models.

The radiative transfer problem is a well-defined problem. It is
described by the radiative transfer equation (RTE), which requires as
input a precise knowledge of the optical properties of the dust
grains, the opacity (the total amount and the spatial distribution of
the dust) and the emissivity (the spatial and energy distribution of
the stars). If a convenient algorithm is found to solve the equation,
the output is the projected image on the sky, or the surface
brightness. But usually the problem is the reverse. Given the observed
image on the plane of the sky, can we recover the three-dimensional
distribution of stars and dust ? The most straightforward way to solve
this problem is to construct a large set of models with various
parameters, solve the transfer equation for each of these models, and
then determine the parameters such that the obtained solution fits the
observations. It is clear that efficient algorithms to solve the RTE
are necessary.

Unfortunately, the RTE is a fairly complicated equation, and it is not
straightforward to solve it, unless some simplifying assumptions are
made. The most widely adopted way to simplify the RTE is an
approximation or downright neglect of the scattering by dust grains
(Guiderdoni \& Rocca-Volmerange~1987, Disney et al.\ 1989, Calzetti et
al.\ 1994, Ohta \& Kodaira~1995). Nevertheless many efforts have been
made to develop methods to solve the RTE exactly. Four of the methods
that have been explored are
\begin{itemize}
\item {\em the spherical harmonics method}, where all terms in the RTE
are expanded into a series of spherical harmonics, such that the RTE
is replaced by a set of ordinary differential equations. 
\item {\em the discretization method}, a technique borrowed from the
stellar atmospheres theory. In this method integrals are replaced by
sums and the differentials by finite differences, resulting in a set
of vector equations which can be solved iteratively.
\item {\em the iteration method}, where intensity is expanded in a
series of partial intensities. Each of the partial intensities obeys
its own RTE, which can be solved iteratively.
\item {\em Monte Carlo simulations}, where the transfer of photons
through the galaxy is investigated by examining the individual paths
of a large number of photons.
\end{itemize}
Each of these methods will have its advantages and disadvantages. On
the one hand the different approaches have different physical
backgrounds, and the selection of a certain algorithm can be useful to
gain physical insight into the problem. For example, the spherical
harmonics method gives direct access to the moments of the radiation
field, whereas the iteration method yields direct information about
the importance of multiple scattering. On the other hand,
computational and practical considerations can be motives to prefer
one method above another. However, these considerations are often
based on general truths without a quantification. For example, it is
generally accepted that Monte Carlo methods are costly, but how costly
in respect with other methods~?

It is important to be able to estimate how a certain solution method
scores in terms of accuracy, efficiency and flexibility. According to
one's specific interest (e.g.\ providing a tool for statistical
studies of attenuation, or constructing detailed radiative transfer
models for individual galaxies), one of these properties may be more
important than another. Knowing the relative performance of the
different methods then allows to select the most suitable candidate
for the problem.

The aim of this paper is to present an unbiased comparison of the four
methods considered. All of them have already been applied to construct
radiative transfer models in order to investigate the attenuation in
disc galaxies (references will be given). However, the assumptions
concerning the geometry and the dust properties made by the various
authors often differ significantly. This makes a comparison of the
various adopted methods in the literature in terms of accuracy,
efficiency and flexibility next to impossible. Therefore we adapt them
such that they can be used to solve the same radiative transfer
problem. We restrict ourselves to a one-dimensional plane-parallel
geometry, however with absorption and multiple scattering properly
accounted for. Moreover, all of these models allow an arbitrary
vertical distribution of stars and dust, and an arbitrary angular
redistribution function (ARF), such that they can serve as a first
approximation to model large portions of galactic discs.

In Section~2 we discuss the radiative transfer problem and present a
disc galaxy model to illustrate the solution techniques, which are
presented in the next four sections. In Section~3 the RTE is solved by
an expansion in spherical harmonics, in Section~4 we describe the
discretization technique, in Section~5 the RTE is solved by an
iterative method, and in Section~6 a Monte Carlo simulation is
used. In Section~7 we compare the different techniques in terms of
accuracy and numerical efficiency, and we discuss them in the light of
their use to construct detailed models for disc galaxies. Section~8
sums up.

\section{The radiative transfer problem}

\subsection{The RTE in plane-parallel geometry}

In plane-parallel geometry, there are only two independent variables
necessary to determine a position and a direction in the galaxy: the
depth $z$, and $\mu$, the cosine of the angle between that direction
and the face-on direction (plane-parallel geometry implies an
azimuthal symmetry around the face-on direction $\mu=1$). The general
form of the transfer equation in plane-parallel geometry can be
written as (in this paper we omit all explicit reference to the
wavelength dependences)
\begin{multline}
	\mu\,\frac{\partial I}{\partial z}(z,\mu)
	=
	-\kappa(z)I(z,\mu)
	+\eta_*(z) \\
	+\frac{1}{2}\omega\kappa(z)
	\int_{-1}^1I(z,\mu')\Psi(\mu,\mu')\txd\mu'.
\label{tv1}
\end{multline}
Here $I(z,\mu)$ is the specific intensity of the radiation,
$\kappa(z)$ is the dust opacity, $\eta_*(z)$ is the stellar
emissivity, $\omega$ the dust albedo and $\Psi(\mu,\mu')$ is the
ARF. The last term in equation~(\ref{tv1}) represents the scattering
term, i.e.\ the net amount of photons which are scattered off dust
grains into the direction $\mu$ from all other directions $\mu'$ at
$z$ (see Appendix~A). We assume that the scattering is coherent, i.e.\
there is no energy redistribution of the scattered light. The thermal
emission of the absorbed light by the dust grains is neglected,
because this occurs at far-infrared wavelengths, whereas we will focus
on shorter wavelengths. Furthermore we assume that the dust grains
have the same properties over the galaxy. This means that the albedo
and the redistribution function are independent of position, and that
both the emissivity and the opacity are separable functions of
position and wavelength.

We replace the height $z$ above the galactic plane by the optical
depth $\tau$, defined as
\begin{equation}
	\tau(z) 
	= 
	\int_z^\infty \kappa(z')\txd z',
\end{equation}
and the RTE then becomes
\begin{equation}
	\mu\,\frac{\partial I}{\partial\tau}(\tau,\mu)
	=
	I(\tau,\mu)
	-S_*(\tau)
	-\frac{\omega}{2}
	\int_{-1}^1I(\tau,\mu')\Psi(\mu,\mu')\txd\mu',
\label{tv2}
\end{equation}
where $S_*(\tau) = \eta_*(\tau)/\kappa(\tau)$ is the stellar source
function. This equation is to be solved for $0\leq\tau\leq\tau_0$, the
total optical depth of the slab,
\begin{equation}
	\tau_0
	= 
	\int_{-\infty}^\infty \kappa(z)\txd z.
\label{deftau0}
\end{equation}
The boundary conditions for our radiative transfer problem are
\begin{equation}
	I(0,-\mu) = I(\tau_0,\mu) = 0 
	\qquad\text{for $\mu>0$,}
\label{bounds}
\end{equation}
which means that there is no incident radiation on either side of the
galaxy. The RTE~(\ref{tv2}), together with the boundary
conditions~(\ref{bounds}) allow us to calculate the radiation field at
any position and into any direction in the galaxy. In the first place
however, we are interested in the radiation field $I(0,\mu)$ that
leaves the galaxy at $\tau=0$ into a certain direction $\mu$, and more
precisely in the fraction of this intensity that is
attenuated. Because the RTE is a linear equation, this fraction will
be independent of the total amount of stellar emission. We can thus
choose the normalization of the stellar emissivity (or the source
function) arbitrarily. We take
\begin{equation}
	\int_{-\infty}^{\infty}\eta_*(z)\,\txd z
	=
	\int_0^{\tau_0}S_*(\tau)\,\txd\tau
	=
	1,
\label{normeta}
\end{equation}
which means that the intensity that leaves the galaxy in the absence
of dust in the face-on direction equals 1. In another direction $\mu$,
the dust-free intensity is then simply $1/\mu$ and the attenuation (in
magnitudes) is
\begin{equation}
	A(\mu) 
	= 
	-2.5\log\left[\mu\,I(0,\mu)\right].
\label{defA}
\end{equation}

\subsection{A disc galaxy model}

In order to solve the problem we still need to characterize the
parameters and functions that appear in~(\ref{tv1}), i.e.\ the optical
properties of the dust and the spatial distribution of stars and dust
have to be specified. Because the present paper focuses on the
comparison of different methods to solve the RTE rather than on an
investigation of the impact of dust attenuation on disc galaxies, we
restrict ourselves here to a single galaxy model to illustrate our
results. For a thorough study of the attenuation curve as a function
of various parameters determining the distribution of stars and dust,
we refer to the second paper in this series (Baes \& Dejonghe~2001b).

The vertical distribution of stars in disc galaxies is still a matter
of debate. Stars were first believed to be isothermally distributed in
the $z$-direction (van der Kruit \& Searle 1981), and later on an
exponential behavior was proposed (Wainscoat et al.\ 1989). Nowadays
it is believed that the true distribution lies in between these two
profiles (Pohlen et al.\ 2000, Schwarzkopf \& Dettmar~2000). We adopt
an exponential profile,
\begin{gather}
	\eta_*(z)
	=
	\frac{1}{2h}\,{\text{e}}^{-|z|/h},
\label{eta}
\end{gather}
where $h$ is the scaleheight, which satisfies the normalization
condition~(\ref{normeta}). For the dust distribution we also assume an
exponential distribution. In a normal disc galaxy, the interstellar
matter will sink down to the central plane and form an obscuring layer
which is narrower than the stellar layer. Detailed radiative transfer
models of edge-on galaxies indicate that the scaleheight of the dust
is about half of the stellar one (Xilouris et al.\ 1999). As opacity
function we then obtain
\begin{equation}
	\kappa(z) 
	=	
	\frac{\tau_0}{h}\,{\text{e}}^{-2|z|/h}.
\label{kappa}
\end{equation}
The normalization of~(\ref{kappa}) is such that~(\ref{deftau0}) is
satisfied. For the total optical depth we adopt a moderate value
$\tau_V=1$, which seems to be appropriate for disc galaxies, as we
argued in the introduction.


\begin{table}
\centering
\caption{The adopted values for the optical properties of the dust
grains. Tabulated are the optical depth $\tau$ relative to the
$V$-band value, the scattering albedo $\omega$ and the asymmetry
parameter $g$.}
\label{optprop.tab}
\begin{tabular}{ccccc} \hline 
band & $\lambda$\,($\mu$m) & $\tau$ & $\omega$ & $g$ \\ \hline
$U$ & 360 & 1.60 & 0.57 & 0.49 \\
$B$ & 440 & 1.32 & 0.57 & 0.48 \\
$V$ & 550 & 1.00 & 0.57 & 0.44 \\
$R$ & 700 & 0.73 & 0.54 & 0.37 \\
$I$ & 850 & 0.47 & 0.51 & 0.31 \\ \hline 
\end{tabular}
\end{table}


For $\Psi(\mu,\mu')$ we adopt the Henyey-Greenstein ARF (see
Appendix~A), appropriate to describe anisotropic scattering. It is a
one-parameter function with a (wavelength dependent) parameter $g$,
which is called the asymmetry parameter and is the average of the
cosine of the scattering angle. Values for the optical properties of
the dust grains (the wavelength dependence of the opacity $\kappa$,
the albedo $\omega$ and the asymmetry parameter $g$) are those
calculated by Maccioni \& Perinotto~(1994) and displayed in Di
Bartolomeo et al.\ (1995). The adopted values of the $U$, $B$, $V$,
$R$ and $I$ bands are tabulated in Table~{\ref{optprop.tab}}.

We wish to stress again that this galaxy model is only illustrative,
and that the methods discussed in this paper are applicable to {\em
any} galaxy model of the kind we consider here, i.e.\ with arbitrary
vertical distribution of stars and dust and with arbitrary ARF.

\section{The spherical harmonics method}

One of the most popular techniques to solve the radiative transfer
equation is a method which uses an expansion in spherical
harmonics. Many papers have been written which describe this method in
detail, e.g.\ Davison~(1957), Flannery et al.\ (1980), Roberge~(1983),
Di Bartolomeo et al.\ (1995). Our approach is based on Roberge~(1983),
whose method can account for any vertical distribution of stars and
dust, and uses the adaptations of Di Bartolomeo et al.\ (1995) in
order to make the matrix in the eigenvalue problem symmetric.

In plane-parallel symmetry, spherical harmonics reduce to Legendre
polynomials, and we expand the intensity and the redistribution
function in a series of Legendre polynomials,
\begin{gather}
	I(\tau,\mu) 
	= 
	\sum_{l=0}^\infty (2l+1)f_l(\tau)P_l(\mu) 
	\label{exp1}
	\\
	\Psi(\mu,\mu') 
	= 
	\sum_{l=0}^\infty (2l+1)\sigma_lP_l(\mu)P_l(\mu').
	\label{exp2}
\end{gather}
The coefficients $f_l(\tau)$ are unknown functions, whereas the
coefficients $\sigma_l$ are known. For example, in the case of
Henyey-Greenstein scattering, $\sigma_l=g^l$, with $g$ the asymmetry
parameter (Appendix~A). Inserting the expansions~(\ref{exp1})
and~(\ref{exp2}) in~(\ref{tv2}), and using the recurrence relations
for Legendre polynomials, we find
\begin{equation}
	\sum_{l=0}^\infty
	\Bigl[
	lf_{l-1}'(\tau)
	+(l+1)f_{l+1}'(\tau)
	-h_lf_l(\tau)
	\Bigr]P_l(\mu)
	=
	-S_*(\tau),
\end{equation}
where we set $h_l=(2l+1)(1-\omega\sigma_l)$. Defining the functions
$\psi_l$ and $g_l$ as
\begin{gather}
	\psi_l(\tau)
	=
	\sqrt{h_l}f_l(\tau)
	\\
	g_l(\tau) 
	=
	-\frac{S_*(\tau)}{\sqrt{h_0}}\delta_{l,0},
\end{gather}
we find an infinite set of linear first-order differential equations
\begin{equation}
	\frac{l}{\sqrt{h_lh_{l-1}}}\psi_{l-1}'(\tau)
	+
	\frac{l+1}{\sqrt{h_{l+1}h_l}}\psi_{l+1}'(\tau)
	=
	\psi_l(\tau) + g_l(\tau).
\end{equation}
We adopt the so-called $P_L$ approximation (consists in assuming
$\psi_l(\tau)=0$ for $l>L$, for a certain value of $L$, $L$ odd) in
order to turn this infinite set into a finite one, which we can write
as a vector equation
\begin{equation}
	{\mathcal A}\boldsymbol{\psi}'(\tau) 
	= 
	\boldsymbol{\psi}(\tau) + \boldsymbol{g}(\tau).
\label{vecrel}
\end{equation}
The RTE is thus reduced to a set of ordinary differential equations,
for which the system matrix ${\mathcal A}$ is a non-singular,
symmetric, tridiagonal matrix with fixed coefficients. Such a problem
is best solved by diagonalization of the matrix, i.e.\ an eigenvalue
problem. The procedure to obtain the expressions for $\psi_l(\tau)$,
are nicely described in Roberge~(1983).

We note in particular that for the $P_L$ solution, the two boundary
conditions~(\ref{bounds}) can only be satisfied for $(L+1)/2$
directions $\mu_i$, which are the positive zeros of the Legendre
polynomial of order $L+1$. The intensity
\begin{equation}
	I(\tau,\mu)
	=
	\sum_{l=0}^L
	\frac{2l+1}{\sqrt{h_l}}\,\psi_l(\tau)\,P_l(\mu).
\end{equation}
at other directions can be obtained by a cubic spline
interpolation. In general we found that the $P_L$ solution converges
very fast, and that the boundary conditions were met when
$L\gtrsim15$, which we adopted in our calculations. This value is
slightly larger than that used by Di Bartolomeo et al.\ (1995).

\section{The discretization method}

One of the first major efforts to investigate the effects of
absorption and multiple scattering in disc galaxies, is the paper by
Bruzual et al.\ (1988). They solve the RTE for a galaxy model
consisting of a homogeneous mixture of stars and dust. Their technique
is based on a discretization of the RTE on a fixed mesh of points, and
difference equations replace the differential equations. We extend the
technique of Bruzual et al.\ (1988) such that it can account for any
vertical distribution of stars and dust, i.e.\ any source function
$S_*(\tau)$. As we will discuss in section~{\ref{dsf}}, we will have
to make a distinction between source functions that remain finite
everywhere and source functions that diverge in places.

\subsection{Finite source functions}

If the source function $S_*(\tau)$ remains finite for all values of
$\tau$, we can easily generalize the procedure of Bruzual et al.\
(1988). Starting from the transfer equation~(\ref{tv2}) we introduce
the even and odd fields
\begin{gather}
	u(\tau,\mu) 
	= 
	\tfrac{1}{2}\left[I(\tau,\mu)+I(\tau,-\mu)\right]
	\\
	v(\tau,\mu) 
	= 
	\tfrac{1}{2}\left[I(\tau,\mu)-I(\tau,-\mu)\right].
\label{tvu}
\end{gather}
The transfer equation can then be written as
\begin{gather}
	\mu\,\frac{\partial u}{\partial\tau}(\tau,\mu)
	=
	v(\tau,\mu)
	-\omega
	\int_0^1v(\tau,\mu')\Psi^o(\mu,\mu')\txd\mu'
	\\
	\mu\,\frac{\partial v}{\partial\tau}(\tau,\mu)
	=
	u(\tau,\mu) - S_*(\tau)
	-\omega\int_0^1u(\tau,\mu')\Psi^e(\mu,\mu')\txd\mu',
\label{tvv}
\end{gather}
where
\begin{gather}
	\Psi^e(\mu,\mu') 
	=
	\tfrac{1}{2}\left[\Psi(\mu,\mu')+\Psi(\mu,-\mu')\right]
	\\	
	\Psi^o(\mu,\mu') 
	=
	\tfrac{1}{2}\left[\Psi(\mu,\mu')-\Psi(\mu,-\mu')\right].
\end{gather}
The new boundary conditions are
\begin{gather}
	u(0,\mu) = v(0,\mu) 
	\label{bound2a}
	\\
	u(\tau_0,\mu) = -v(\tau_0,\mu).
	\label{bound2b}
\end{gather}
These equations are analogous to the ones Bruzual et al.\ (1988) use
for their homogeneous slab, with the exception that the constant
source function for the homogeneous slab is replaced by a
non-constant, but finite source function $S_*(\tau)$. A similar
discretization procedure can be followed. We introduce a grid of $N+1$
mesh points $\tau_i$, uniformly spaced in optical depth space. In
between these points we introduce a second grid, denoted by
half-integer numbers $\tau_{i+1/2}$. A derivative evaluated in a
half-integer grid point $\tau_{i+1/2}$ becomes a difference evaluated
in the nearest integer mesh points $\tau_i$ and $\tau_{i+1}$, and
analogous for the integer mesh points. At the boundaries $\tau_0$ and
$\tau_N$ the derivatives are expressed by means of the boundary
conditions. The integrals over $\mu$ are approximated by $M$-point
Gauss-Legendre quadrature, and we hence introduce a mesh of $M$ points
$\mu_j$, being the roots of the $M$th order Legendre polynomial. The
RTEs are then replaced by a set of linear vector equations, where the
$(2N+1)M$ unknowns are the even and odd fields $u_{i,j}$ and
$v_{i+1/2,j}$ evaluated at the integer and half-integer mesh points
respectively. These equations are solved recursively using (a
simplification of) the elimination scheme of Milkey et al.\ (1975).

Optimal values for $M$ and $N$ are hard to determine. On the one hand
these values should not be too high, because of computational
limitations. Indeed, the elimination scheme consists of two loops, and
every step in the first loop consists of 2 matrix inversions, where
the order of the matrices is $M$, the number of angle quadrature
points. This means that totally $2N+1$ matrices of order $M$ need to
be inverted, which is a numerically costly process. Moreover, the
memory requirement is high, because the $2N+1$ matrices calculated in
the first loop need to be stored, for use in the second loop. On the
other hand, the values for $M$ and $N$ should not be taken too low,
such that the approximation of differentials by differences and
integrals by quadrature sums is acceptable. Typically $M=10$, while
typical values for $N$ are dependent on the total optical depth
$\tau_0$ and should be chosen such that $\Delta\tau$ never exceeds
0.01. For our galaxy model this meant $N\gtrsim100$.

\subsection{Diverging source functions}
\label{dsf}

The above procedure cannot be used anymore when the source function
diverges at the boundaries $\tau=0$ and/or $\tau=\tau_0$, because the
source function needs to be evaluated in these end points. Such source
functions are realistic, because they correspond to all distributions
where the dust distribution is narrower than the stellar distribution,
which is observed in most galaxies. In particular, the galaxy model
described in Section~2.2 has a diverging source function.

One obvious solution would be to apply a cutoff, but the method then
becomes very dependent on the actual position of the cutoff. Moreover
such a solution is inaccurate, because it involves a lot of matrix
inversions, and these are difficult operations if the matrices contain
elements of very different magnitudes.

If the source function diverges, we don't use the optical depth
coordinates, but we define a similar independent variable $\xi$,
\begin{equation}
	\xi(z) 
	= 
	\int_z^\infty\eta_*(z')\,\txd z',
\end{equation}
which assumes values between 0 and 1, because of the normalization of
the emissivity~(\ref{normeta}). The RTE~(\ref{tv1}) becomes
\begin{multline}
	\mu\,\frac{\partial I}{\partial\xi}(\xi,\mu)
	=
	R_*(\xi)\,I(\xi,\mu)
	-1
	\\
	-\frac{1}{2}\omega R_*(\xi)
	\int_{-1}^1I(\xi,\mu')\Psi(\mu,\mu')\txd\mu',
\label{tv3}
\end{multline}
where $R_*$ is the reciprocal of the stellar source function,
\begin{equation}
	R_*(\xi) 
	= 
	\frac{1}{S_*(\xi)}
	=
	\frac{\kappa(\xi)}{\eta_*(\xi)}.
\end{equation}
Starting from this form of the RTE, we can repeat the discretization
procedure from the previous paragraph, i.e.\ we construct a uniform
mesh $\xi_i$ and an intermediate mesh $\xi_{i+1/2}$ for the finite
differences, and a mesh $\mu_j$ for the Legendre-Gauss quadrature, and
we obtain a set of vector equations for the unknowns $u_{i,j}$ and
$v_{i+1/2,j}$, which are solved by the elimination scheme of Milkey et
al.\ (1975).

The arguments leading to the choice of $N$ and $M$ are the same as in
the previous paragraph.

\section{Iteration methods}

\subsection{The intensity as a series}

In a very early paper on scattering, Henyey~(1937) shows that the
RTE~(\ref{tv2}) can be solved iteratively by writing the intensity as
a summation of partial intensities $I_n$, which represent the
radiation field consisting of photons that have been scattered exactly
$n$ times. This technique has been adopted by van de Hulst \& de
Jong~(1969) and Xu \& Helou~(1996) to solve the RTE in plane-parallel
geometry. Specifically in our case, we write the intensity as
\begin{equation}
	I(\tau,\mu)
	=
	\sum_{n=0}^\infty I_n(\tau,\mu),
\label{Ialssom}
\end{equation}
The $n$th partial intensity satisfies the RTE
\begin{equation}
	\mu\,\frac{\partial I_n}{\partial\tau}(\tau,\mu)
	=
	I_n(\tau,\mu)
	-S_n(\tau,\mu),
\label{partialRTE}
\end{equation}
where
\begin{gather}
	S_0(\tau,\mu) 
	=
	S_*(\tau) 
	\\
	S_n(\tau,\mu) 
	=
	\frac{\omega}{2}
	\int_{-1}^1I_{n-1}(\tau,\mu')\Psi(\mu,\mu')\txd\mu',
\label{intang}
\end{gather}
and is subject to the boundary conditions
\begin{equation}
	I_n(0,-\mu) = I_n(\tau_0,\mu) = 0
	\qquad\text{for $\mu>0$.}
\end{equation}	
The solution of~(\ref{partialRTE}) can be directly written
\begin{gather}
	I_n(\tau,\mu) 
	=
	\frac{1}{\mu}
	\int_{\tau}^{\tau_0}
	S_n(\tau',\mu)\,
	\exp\left(\frac{\tau-\tau'}{\mu}\right)
	\txd\tau'
	\\
	I_n(\tau,-\mu) 
	=
	\frac{1}{\mu}
	\int_0^{\tau}
	S_n(\tau',\mu)\,
	\exp\left(\frac{\tau'-\tau}{\mu}\right)
	\txd\tau'.
\end{gather}
We calculated the partial intensities and source functions on a mesh
of size $L$ in optical depth and size $M$ in angle. Because the
integration over the angle in~(\ref{intang}) always has the same
integration domain, we chose the $M$ angle points as the abscissae for
an $M$-point Gauss-Legendre quadrature. For the calculation of the
$I_n$, the integration along the path has variable boundaries, and
hence a simple quadrature would yield bad results near the edges of
the mesh. Instead, we used a uniform mesh of optical depth points, and
for each fixed angle $\mu_j$, we constructed a cubic spline
approximation to $S_n(\tau,\mu_j)$, and performed the integration
using this function. Typical values for $L$ and $M$ are 40 and 10
respectively.

The number of terms necessary in the summation~(\ref{Ialssom}) in
order to obtain an accurate result depends on the wavelength, because
the $n$th term in the expansion is proportional to the $n$th power of
the albedo $\omega^n$. Xu \& Helou~(1996) used 20~terms for their
sandwich model; we find that in general, about 10~terms are
sufficient. For our one-dimensional plane-parallel geometry this
number of integrations is still manageable. However, if the method is
extended to more complicated geometries, more dimensions will be
added, and calculation of such a high number of terms becomes very
time-consuming, such that adaptations of the method are desirable.

\subsection{The intensity as a geometric series}


\begin{figure}
\centering 
\includegraphics[clip,bb=186 467 408 625]{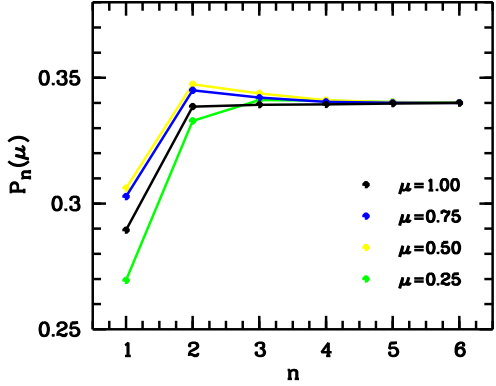}
\caption{The ratio $P_n(\mu) = I_n(0,\mu)/I_{n-1}(0,\mu)$ at the $V$
band of two consecutive terms in the series~(\ref{Ialssom}), as a
function of $n$. It is shown for four different inclination angles,
for the model described in Section~2.2.}
\label{kylbah.ps}
\end{figure}


A simplification of the iteration method was introduced by Kylafis \&
Bahcall~(1987). They use the approximation that the amount of photons
that are scattered exactly $n+1$ times to the amount of photons
scattered exactly $n$ times, is constant, i.e.
\begin{equation}
	\frac{I_{n+1}(\tau,\mu)}
	{I_n(\tau,\mu)}
	=
	\frac{I_n(\tau,\mu)}
	{I_{n-1}(\tau,\mu)}
	\qquad\text{for all $n$.}
\label{KBcond}
\end{equation}
If this approximation holds, only two terms in the sum~(\ref{Ialssom})
need to be calculated, because it can be written as a geometric series
\begin{equation}
	I(\tau,\mu)
	=
	I_0(\tau,\mu)
	\sum_{j=0}^{\infty}
	\left[
	\frac{I_1(\tau,\mu)}{I_0(\tau,\mu)}
	\right]^j.
\end{equation}
To test the accuracy of the approximation~(\ref{KBcond}), we
calculated several terms in the sum~(\ref{Ialssom}) for our galaxy
model. The results are shown in Figure~{\ref{kylbah.ps}}, where we
plot the ratio $P_n(\mu) = I_n(0,\mu)/I_{n-1}(0,\mu)$ as a function of
$n$, for four different values of $\mu$. It is clear that $P_n(\mu)$
is indeed constant for $n\gtrsim3$, and that this constant is
independent of the angle $\mu$. For the first values of $n$ however,
the condition~(\ref{KBcond}) is clearly not satisfied. Therefore we
adapt the strategy of Kylafis \& Bahcall~(1987), and we assume that
the condition~(\ref{KBcond}) is valid for $n>N$. The intensity can
then be written as
\begin{multline}
	I(\tau,\mu)
	=
	I_0(\tau,\mu)
	+
	I_1(\tau,\mu)
	+
	\cdots
	\\+
	I_{N-1}(\tau,\mu)
	\sum_{j=0}^{\infty}
	\left[
	\frac{I_N(\tau,\mu)}{I_{N-1}(\tau,\mu)}
	\right]^j.
\label{KB87alg}
\end{multline}
The parameter $N$ thus determines the last of the partial intensities
that needs to be calculated explicitly. We find that already for $N=2$
the results are correct with less than a one-percent error. The number
of integrations that then has to be performed for the solution of the
RTE is~$3LM$.

\section{Monte Carlo simulation}

The last method we considered to solve the RTE is a Monte Carlo
simulation, which is probably the most widely adopted method for
radiative transfer problems. The principles of this method are
described in detail by Cashwell \& Everett~(1959), Mattila~(1970),
Witt~(1977), Yusef-Zadeh et al.\ (1984) and Bianchi et al.\ (1996).

\subsection{The principles}

The Monte Carlo method basically follows the individual path of a very
large number $N$ of photons through the galaxy. At each moment, the
fate of a photon on its path is determined by a number of quantities
such as the free path length between two interactions, the nature of
the interaction (scattering or absorption) and the direction change
during a scattering event. Each of these quantities can be described
by a random variable, taken from a particular probability density
$p(x)\txd x$.

More specifically, in plane-parallel geometry a photon is
characterized by two variables: the position (or equivalently the
optical depth $\tau$) and the direction $\mu$. To start, the initial
position $\tau_1$ and direction $\mu_1$ are generated randomly as
\begin{gather}
	X_1 = \int_0^{\tau_1} S_*(x)\,\txd x \\
	X_2 = \int_{-1}^{\mu_1}\frac{\txd x}{2}
	\quad\text{or}\quad
	\mu_1 = 2X_2-1,
\end{gather}
where $X_1$ and $X_2$ are uniform deviates. Next we generate a random
free path length $\ell$ (also in optical depth units) by setting
\begin{equation}
	X_3 = \int_0^{\ell}{\text{e}}^{-x}\txd x
	\quad\text{or}\quad
	\ell = -\ln(1-X_3)
\label{x3}
\end{equation}
This randomly determined path length is to be compared with the
maximal free path length $L$ of the photon in consideration,
\begin{equation}
	L
	=
	\begin{cases}
	\,\,\dfrac{\tau_1}{\mu_1} & \quad\text{if $\mu_1>0$} \\
	\,\,-\dfrac{\tau_0-\tau_1}{\mu_1} & \quad\text{if $\mu_1<0$}.
	\end{cases}
\end{equation}
If $\ell>L$, the photon will leave the galaxy, and its direction
$\mu_1$ is recorded. If $\ell<L$, the photon will interact with a dust
grain. The nature of this interaction can be determined by chosing a
uniform deviate $X_4$ and setting
\begin{equation}
	\text{interaction}
	=
	\begin{cases} 
	\,\,\text{scattering} & \qquad\text{if $X_4<\omega$} \\
	\,\,\text{absorption} & \qquad\text{if $X_4>\omega$}.
	\end{cases}
\end{equation}
If the interaction is an absorption, the photon disappears and will
not contribute to the final intensity. If the interaction is a
scattering, the photon will have a new position and a new
direction. Given the free path length $\ell$ and the original position
$\tau_1$ and direction $\mu_1$, the new position of the photon is
\begin{equation}
	\tau_2 = \tau_1-\mu_1\ell,
\end{equation}
whereas the new direction $\mu_2$ is determined by the uniform deviate
\begin{equation}
	X_5 = \frac{1}{2}\int_0^{\mu_2}\Psi(\mu_1,x)\,\txd x.
\end{equation}
This procedure can now be repeated until the photon is either absorbed
or leaves the galaxy. In the latter case it will contribute to the
observed intensity and its direction must be recorded. After a
sufficiently large number of such experiments, an $M$-binned histogram
of the emerging angular distribution can be constructed. Due to the
planar symmetry of our galaxy models, photons leaving the galaxy at
the back side can be added to those leaving at the front side. The
ratio of the number of photons to the number of photons that would be
in the bin if there were no dust attenuation (i.e.\ the distribution
of the initial directions), then yields the attenuation in that
bin. The final attenuation curve $A(\mu)$ is determined by fitting a
smooth curve to these results.

\subsection{Optimizing the routine}

There are two simple ways in which the above described routine can be
optimized, i.e.\ adapted such that better statistics can be obtained
with less photons.

The first adaptation is to avoid that photons disappear from the
radiation field by absorption. This can be done by assigning a weight
to each photon. At each interaction the probability that the photon
will be scattered is $\omega$, whereas the probability for absorption
is $1-\omega$. Instead of simulating absorption as described above, we
let the photon scatter and we reduce its weight by a factor
$\omega$. This way, no photons disappear from the radiation field, and
each photon will eventually leave the galaxy. At that point, both
direction and weight are recorded. The final histogram is obtained by
counting the number of photons in each bin, weighted by their
individual weight.

A second adaptation is the concept of forced first scattering, by
which each photon is forced to be scattered at least once before it
leaves the galaxy. Given the initial maximal free path length
$L(\tau,\mu)$ of the photon, the probability that the photon leaves
the galaxy is $\exp(-L)$. When the total optical depth of the galaxy
is low, many photons leave the galaxy without interaction, such that a
large number of photons is necessary to obtain reliable statistics of
the scattered radiation. Hence, instead of allowing that a photon can
escape from the galaxy, we split it. A fraction with weight $w =
\exp(-L)$ leaves the galaxy and will be classified. The other fraction
with weight $w = \omega[1-\exp(-L)]$ is forced to scatter before it
leaves the galaxy. Therefore a free path length $\ell$ has to be
generated such that $\ell<L$. This can be done by replacing~(\ref{x3})
with
\begin{multline}
	X_3 
	= 
	\int_0^{\ell}{\text{e}}^{-x}\txd x \left/
	\int_0^{L}{\text{e}}^{-x}\txd x \right.
	\\
	\quad\text{or}\quad
	\ell
	=
	-\ln\left[1-X_3\left(1-{\text{e}}^{-L}\right)\right].
\end{multline}
After this forced first scattering, the following scatterings are as
described above.

These two concepts reduce the number of photons, necessary to obtain
reliable statistics, significantly. Typical values of $N$ and $M$ we
use in our calculations are $N=10^5$ and $M=100$, such that the mean
number of photons in each bin is around 1000.

\section{Discussion of the adopted techniques}


\begin{figure*}
\centering 
\includegraphics[clip,bb=110 461 484 748]{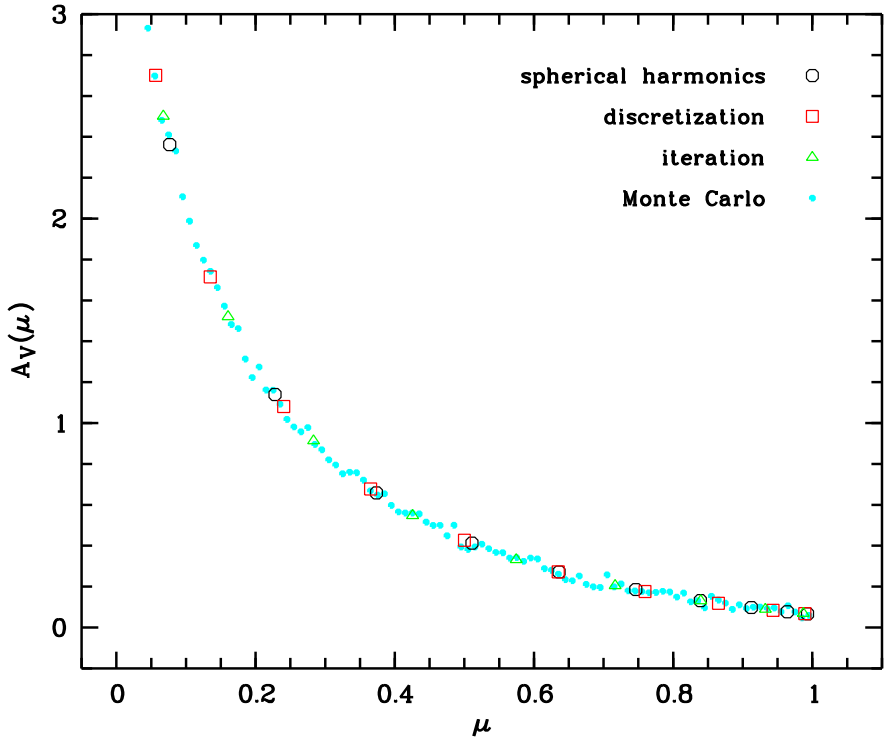}
\caption{Comparison of the $V$ band attenuation curve $A_V(\mu)$
obtained by the four different methods. The results are only shown for
the particular values for which the different methods yield the
solution directly (see text).}
\label{accu.ps}
\end{figure*}


\begin{table}
\centering
\caption{A check on the accuracy of the four adopted methods of
solving the RTE. We computed the attenuation $A(\mu)$ of the model
described in Section~2.2, for four angles and for the five bands $U$,
$B$, $V$, $R$ and $I$. The four results correspond to the spherical
harmonics method ({\em sh}), the discretization method ({\em di}), the
iteration method ({\em it}) and the Monte Carlo simulation ({\em
mc}).}
\begin{tabular}{cccccc} \hline
\label{accu.tab}
& $\mu$ & $0.25$ & $0.50$ & $0.75$ & $1.00$ \\ \hline
$U$ & sh & 1.113 & 0.617 & 0.356 & 0.209 \\
    & di & 1.113 & 0.617 & 0.355 & 0.209 \\
    & it & 1.119 & 0.617 & 0.354 & 0.208 \\
    & mc & 1.108 & 0.616 & 0.360 & 0.213 \\[1mm]
$B$ & sh & 1.010 & 0.520 & 0.278 & 0.149 \\
    & di & 1.010 & 0.520 & 0.277 & 0.148 \\
    & it & 1.014 & 0.519 & 0.276 & 0.148 \\
    & mc & 1.003 & 0.524 & 0.277 & 0.129 \\[1mm]
$V$ & sh & 0.865 & 0.398 & 0.187 & 0.081 \\
    & di & 0.866 & 0.398 & 0.187 & 0.081 \\
    & it & 0.867 & 0.397 & 0.186 & 0.081 \\
    & mc & 0.851 & 0.402 & 0.188 & 0.078 \\[1mm]
$R$ & sh & 0.721 & 0.299 & 0.123 & 0.038 \\
    & di & 0.721 & 0.298 & 0.123 & 0.037 \\
    & it & 0.722 & 0.298 & 0.123 & 0.038 \\
    & mc & 0.710 & 0.300 & 0.127 & 0.029 \\[1mm]
$I$ & sh & 0.537 & 0.197 & 0.068 & 0.007 \\
    & di & 0.538 & 0.196 & 0.067 & 0.006 \\
    & it & 0.538 & 0.196 & 0.067 & 0.007 \\
    & mc & 0.533 & 0.195 & 0.070 & 0.007 \\ \hline
\end{tabular}
\end{table}


\subsection{Comparison of the numerical results}

Because we have at our disposal four different methods to solve the
RTE, it is straightforward to check the correctness of the individual
methods by comparing their results. In Figure~{\ref{accu.ps}} we plot
the $V$ band attenuation for the model described in Section~2.2, for
the four methods considered. We can obviously conclude that the four
modelling procedures are in complete agreement with each other.

However, as already mentioned, the modelling techniques do not reveal
the solution at any direction $\mu$. The spherical harmonics method
only yields $A$ for the $(L+1)/2$ positive zeros of the Legendre
polynomial of order $L+1$. The discretization and iteration methods
yield $A$ at the abscissae of the adopted quadrature formulae. The
Monte Carlo method actually yields a histogram of the attenuation in
each of the bins into which the interval of possible $\mu$-values is
divided. In order to find the attenuation in a randomly chosen
direction $\mu$, we use a spline interpolant for the first three
methods, and a fitting polynomial for the Monte Carlo method. In
Table~{\ref{accu.tab}} we tabulate the attenuation curve $A(\mu)$,
calculated by each of the four methods, for four inclination angles
$\mu$. This table shows that the differences between the different
solutions is always of the order $\Delta A
\approx0.01$~mag. Figure~{\ref{accu.ps}} and Table~{\ref{accu.tab}}
prove that the four methods are accurate and consistent.

In principle, we could also check the correctness of the methods by
comparing our results with those obtained by other teams, who adopted
similar techniques. Such a comparative analysis is conducted by Di
Bartolomeo et al.\ (1995), who compared their results for a
homogeneous slab with those of Guiderdoni \& Rocca-Volmerange (1987),
Kylafis \& Bahcall~(1987), Bruzual et al.\ (1988), Witt et al.\
(1992), Byun et al.\ (1994) and Calzetti et al.\ (1994). The
discrepancies between the extinction curves are significant
(e.g. $\Delta A_U$ up to 0.3~mag for $\tau_V$ as small as 0.5), and it
is important to investigate why this is so. Are they due to the
adopted solution technique or to other causes~?  Generally, the
discrepancies can be the result of differences in
\begin{enumerate}
\item {\em the solution method.} In five of the mentioned papers the
RTE is solved exactly (i.e.\ taking absorption and scattering fully
into account) using the discretization, iteration or Monte Carlo
techniques, whereas Guiderdoni \& Rocca-Volmerange~(1987) and Calzetti
et al.\ (1994) use approximate analytical solutions.
\item {\em the grain properties.} The fact that different authors use
different sets of optical properties can introduce substantial
discrepancies in the attenuation curve. These differences between the
various values for the grain properties can be very substantial, as
clearly shown in Figures~1 and~2 of Di Bartolomeo et al.\ (1995).
\item {\em the geometry.} As Di Bartolomeo et al.\ (1995) indicated,
it is sometimes impossible to compare results because the geometrical
configuration used by the various authors is not always the same. Witt
et al.\ (1992) use a spherical symmetry, Kylafis \& Bahcall~(1987) and
Byun et al.\ (1994) adopt a axisymmetric galaxy model, whereas the
other authors use a plane-parallel homogeneous slab.
\end{enumerate}
In fact in only two of the seven papers, the RTE is solved exactly for
a plane-parallel homogeneous slab: Bruzual et al.\ (1988) by using the
discretization technique and Di Bartolomeo et al.\ (1995) by adopting
the expansion in spherical harmonics. However, the authors adopt a
different set of optical properties of the dust grains, and the
differences $\Delta A(\mu)$ between the attenuation curves can still
be due to the first two points mentioned above.

\begin{figure}
\centering 
\includegraphics[clip,bb=175 468 408 683]{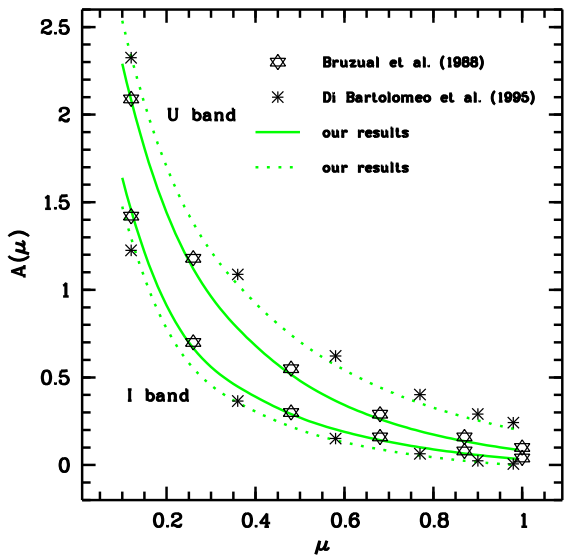}
\caption{Comparison of our work with that of Bruzual et al.\ (1988)
and Di Bartolomeo et al.\ (1995). Shown is the attenuation curve for
the $U$ and $I$ bands, for a homogeneous slab with $\tau_V=1$. The
stars and asterixes represent the results obtained by Bruzual et al.\
(1988) and Di Bartolomeo et al.\ (1995) respectively, the curves are
our results where we used the corresponding dust parameters.}
\label{literature.ps}
\end{figure}

To test to which degree the adopted technique contributes to the
differences in the extinction curve, we consider a plane-parallel
slab, and solve the RTE {\em twice} for each of the four methods at
our disposal: once with the dust grain parameters of Di Bartolomeo et
al.\ (1995) --the ones we adopt throughout this paper--, and once with
those of Bruzual et al.\ (1988). For a galaxy with total optical depth
$\tau_V=1$, the results are shown in Figure~{\ref{literature.ps}} for
the $U$ and $I$ bands. We find a very good agreement between the
results obtained by Bruzual et al.\ (1988) and Di Bartolomeo et al.\
(1995) respectively, and our results with the corresponding dust
parameters. This demonstrates that the attenuation differences $\Delta
A$ are completely due to the differing dust properties, and that the
spherical harmonics and iteration methods are reliable.


\subsection{Computational efficiency}

Besides being accurate, another desirable quality of a solution method
is the computational efficiency, as we explained in the introduction.

For the spherical harmonics method the only numerically costly
operations are the calculation of the eigenvalues of a matrix of order
$L+1$, the inversion of such a matrix, and $L+1$ simple
integrations. Given $L\approx15$ this cost is relatively low. For the
discretization method $2N+1$ matrices of the order $M$ need to be
inverted, with typical values $M=12$ and $N=100$. Because the
computation time for the inversion of a matrix of order $M$ is
proportional to $M^{2.8}$ (Press et al.\ 1989), the numerical cost of
the discretization method will be considerably higher. The iteration
method requires $3M$ cubic spline fits, and $3LM$ integrations along
the line-of-sight, with typical values $L=40$ and $M=10$. Although the
integrands are well-behaved (the product of a smooth source function
with an exponential), such that each individual integration is
relatively easy to perform, the high number of integrations makes the
iteration method numerically costly. Finally, for the Monte Carlo
simulation, the only costly operation for each photon trajectory is
the generation of $2n+3$ pseudo-random numbers, where $n$ is the
number of scatterings for the photon. Because the number of photons
must be fairly high in order to achieve reliable statistics (we use
$N=10^5$), the numerical cost of the Monte Carlo method is
considerable.


\begin{table}
\centering
\caption{A comparison of the computation time necessary for the
calculation of the attenuation curve $A(\mu)$ for a single
wavelength. The number gives the actual computation time in seconds,
the second number is the computation time relative to the spherical
harmonics method.}
\begin{tabular}{ccc} \hline
\label{comptime.tab}
spherical harmonics & 0.091 s & 1 \\
discretization & 2.55 s & 28 \\
iteration & 15.68 s & 172 \\
Monte Carlo & 15.25 s & 168 \\ \hline  
\end{tabular}
\end{table}


In Table~{\ref{comptime.tab}} we tabulate the mean computation time
necessary for the calculation of the attenuation curve for a single
wavelength. This little table shows that the iteration method and the
Monte Carlo simulation have the same efficiency, and that the
spherical harmonics method is significantly more efficient that the
other methods. Although for this simple one-dimensional plane-parallel
geometry the efficiency is less important (the computation times are
very feasible for {\em all} methods), it will be important if we want
to extend these solution to more complex geometries.

\subsection{Extension to more complex geometries}

The implementations described in this paper can handle the RTE in a
plane-parallel geometry, and although they can accomodate any vertical
star-dust geometry, they cannot model real disc galaxies. More
realistic models require a light distribution that also decreases
exponentially in the radial direction (Freeman 1970, Saio \&
Yoshii~1990, Firmani et al.\ 1996). Instead of a one-dimensional
plane-parallel geometry, the RTE then has to be solved in a
two-dimensional axisymmetric geometry, with a set of four independent
coordinates instead of a pair. This extra dimension complicates the
RTE substantially. This observation forces us to think about how the
techniques described in this paper can be generalized to solve the RTE
in axisymmetric disc galaxy models.

The most obvious candidate for extension to axisymmetric galaxy models
(or any other geometry) is the Monte Carlo simulation. This technique
has already been applied several times to realistic disc galaxy models
(e.g.\ Bianchi et al.\ 1996, Ferrara et al.\ 1996, de Jong~1996, Wood
\& Jones~1997, Matthews \& Wood~2000). Besides being able to treat any
geometry, it is also sufficiently flexible to treat, for example,
clumpy dust distributions (Witt \& Gordon~1996, 2000, Bianchi et al.\
2000). This flexibility makes Monte Carlo simulations probably the
most powerful technique for solving complicated radiative transfer
problems. However, their great numerical cost is a disadvantage, and
this cost will grow strongly if the method is extended to axisymmetric
geometries. Indeed, given an initial position, direction and a
pathlength, the next position on the path is directly calculated in
our plane-parallel geometry, but in axisymmetric geometries this
becomes a very time-consuming operation (Bianchi et al.\
1996). Therefore it is worth while to investigate how the other
techniques can be generalized.

Also the iteration technique is easily extendible to more complex
geometries. And in contrast to the Monte Carlo simulation, it can be
expected that the numerical cost will grow in proportion to the number
of dimensions. If meshes of order $J$ and $K$ are constructed inorder
to account for the extra radial and azimuthal dimensions in
axisymmetric geometry, the entire routine can be expected to take
about $JK$ times as much computation times. The iteration method has
been adopted to solve the RTE for axisymmetric disc galaxies in its
original form (Vansevi\v cius et al.\ 1997), but it is the
modification by Kylafis \& Bahcall~(1987) that has increased the
efficiency considerably, and turned the method into a very practical
instrument to investigate dust attenuation in disc galaxies (Bosma et
al.\ 1992, Byun et al.\ 1994, Misiriotis et al.\ 2000). In particular,
the method has been adopted to construct detailed radiative transfer
models for a set of highly inclined disc galaxies (Xilouris et al.\
1997, 1998, 1999), which have been compared with dust emission models
(Alton et al.\ 2000). It is interesting that the results of Xilouris
et al.\ (1999) seem at a first glance not in correspondence with the
Monte Carlo results of Kuchinski et al.\ (1998), whose computed
optical depths are approximately a factor 2-3 higher.

In the previous paragraph, it was shown that, while the efficiency of
the Monte Carlo and iteration techniques is comparable (at least for
the one-dimensional geometry), the spherical harmonics method is
superb in efficiency. It is in fact possible to extend the expansion
in spherical harmonics to geometries other than plane-parallel or
spherical (Davison 1957). However, another way to extend the spherical
harmonics technique to axisymmetric disc galaxies has been proposed by
Corradi et al.\ (1996), who assume a {\em local} plane-parallel
geometry along each line-of-sight. The RTE can then be solved for each
line of sight separately, such that the computation time will be very
small. The limitation of this method however is that the assumption of
a local plane-parallel geometry must remain acceptable. For (nearly)
face-on galaxies this may be the case, because the scale at which the
galaxy's structure (i.e.\ the source function) changes on the plane of
the sky is small compared to the mean free path of the photons. For
highly inclined galaxies however, the mean free path of the photons is
large compared to the scale of variation of the source function on the
plane of the sky, such that the assumption of local plane-parallel
geometry will not be applicable.

The last described method, the discretization method is a typical
one-dimensional technique and is least suitable for extension to more
complex geometries.

\section{Conclusion}

Constructing radiative transfer models for disc galaxies requires a
solution of the RTE. This equation is, even for the most simple
geometries, sufficiently complex, such that sophisticated methods are
necessary to solve them. The RTE typically takes as input the
distribution of stars and dust, and the output is the projected image
on the sky. However, usually the problem is the reverse: the
distribution of dust and stars has to be derived from the observed
intensity. The most simple way to solve this problem is to find a best
fitting solution of the RTE in a large parameter space. This involves
that the RTE has to be solved repeatedly, such that {\em efficiency}
is a very important property of a good solution method. For the same
reason {\em accuracy} is important: the effects of dust on the
attenuation on certain physical parameters can be very weak, such that
a high accuracy is required to determine the stellar and dust
distribution. And last but not least, there are many ways to solve the
RTE for simple one-dimensional geometries, but the modelling of real
galaxies demands methods which are applicable to geometries other than
plane-parallel or spherical, in the first place
axisymmetric. Therefore {\em flexibility} is necessary.

Several techniques are adopted in the literature to solve the RTE,
even for realistic axisymmetric disc galaxy models, but the accuracy,
efficiency and flexibility of these methods has never been properly
compared. Nevertheless, it is clear that knowledge of these properties
is necessary in order to select the most suitable method for a
particular radiative transfer problem.

In this paper we investigate four different methods to solve the RTE
in a simple plane-parallel geometry: the spherical harmonics method,
the discretization method, an iterative method and a Monte Carlo
simulation. All of them are adapted such that they solve the RTE
exactly (i.e.\ with the physical processes of absorption and multiple
scattering properly taken into account), and that they allow an
arbitrary vertical distribution of stars and dust and an arbitrary
ARF. This way our methods can contribute to understanding of the
transfer of radiation in realistic galactic discs.

For a galaxy model with realistic vertical structure and dust
parameters, all four methods yield exactly the same results, with
differences between the attenuation curves at most a few hundredths of
a magnitude. They can thus be considered as accurate. We also compare
our results with those obtained by others, and find that the results
are in agreement with each other. Concerning efficiency, the iteration
method and Monte Carlo method are close, whereas the discretization
method is 6 times as efficient, and the spherical harmonics method
even 170 times. Whereas the Monte Carlo method can easily be
generalized to an arbitrary geometry, we anticipate that the iteration
method will probably be the most efficient routine in axisymmetric
geometry. The adaptation of the routine to axial symmetry is more or
less straightforward, and the efficiency will not suffer as much from
the extra dimensions as the Monte Carlo simulation. This issue will be
thoroughly investigated and presented in a forthcoming paper.

\appendix
\section{The Henyey-Greenstein angular redistribution function}


\begin{figure}
\centering \includegraphics[clip,bb=173 470 421
620]{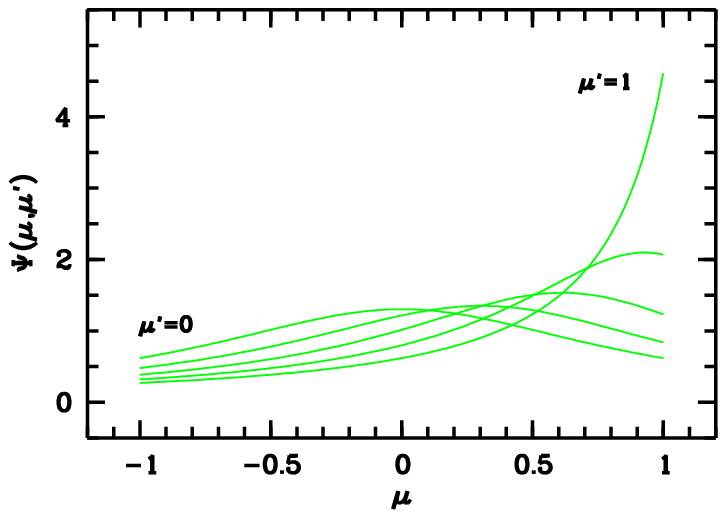}
\caption{The $V$ band Henyey-Greenstein ARF $\Psi(\mu,\mu')$ as a
function of the new direction $\mu$. It is shown for various values of
the initial direction $\mu'$, ranging between 0 and 1, with
intermediate step 0.25. For negative values of $\mu'$ the ARF is
obtained by the symmetry relation $\Psi(\mu,-\mu') =
\Psi(-\mu,\mu')$.}
\label{phasefun.ps}
\end{figure}


In general, the phase function $\Phi(\bfn,\bfn')$ defines the
probability that a photon, that is scattered from a direction $\bfn'$,
will obtain a new direction $\bfn$. If we normalize the phase function
as
\begin{equation}
	\iint\Phi(\bfn,\bfn')\frac{\txd\bfn'}{4\pi} = 1
	\qquad\text{for all $\bfn$},
\end{equation}
then the amount of photons added to the radiation field $I(\bfr,\bfn)$
at a position $\bfr$ into a direction $\bfn$ due to scattering is
\begin{equation}
	\omega\kappa(\bfr)
	\iint I(\bfr,\bfn')\Phi(\bfn,\bfn')
	\frac{\txd\bfn'}{4\pi}.
\end{equation}
It can be assumed that the scattering phase function does not depend
independently on the four variables
$(\bfn,\bfn')=(\mu,\phi,\mu',\phi')$. We assume that it depends on
them only through the angle $\Theta$ between the incident and the
scattered radiation. In our plane-parallel geometry, we have azimuthal
symmetry, and the scattering term can then be written as
\begin{equation}
	\frac{1}{2}\omega\kappa(z)
	\int_{-1}^1I(z,\mu')\Psi(\mu,\mu')\txd\mu',
\end{equation}
where the ARF $\Psi(\mu,\mu')$ is defined as (van de Hulst \& de
Jong~1969, Bruzual et al.\ 1988)
\begin{equation}
	\Psi(\mu,\mu')
	=
	\frac{1}{2\pi}\int_0^{2\pi}\Phi(\cos\Theta)\txd\phi'.
\label{angdist}
\end{equation}
A widely used phase function is the one named after Henyey and
Greenstein~(1941), 
\begin{equation}
	\Phi(\cos\Theta)
	=
	\dfrac{1-g^2}{\left(1+g^2-2g\cos\Theta\right)^{3/2}},
\label{hg}
\end{equation}
which is an accurate one-parameter family to describe the average
scattering in galactic dust. Its parameter $g$ is the asymmetry
parameter and is the mean cosine of the scattering angle,
$g=\langle\cos\Theta\rangle$. A very useful characteristic of this
function is that it has a very simple expansion in Legendre
polynomials. Using the generating function for Legendre polynomials it
is easily shown that
\begin{equation}
	\Phi(\cos\Theta)
	=
	\sum_{l=0}^\infty(2l+1)g^lP_l(\cos\Theta),
\end{equation}
such that we immediately have an expression for the angular
redistribution function (Chandrasekhar~1960, Roberge~1983)
\begin{equation}
	\Psi(\mu,\mu')
	=
	\sum_{l=0}^\infty(2l+1)g^lP_l(\mu)P_l(\mu').
\end{equation}
In order to derive a closed expression for $\Psi(\mu,\mu')$ we use
\begin{equation}
	\cos\Theta 
	= 
	\mu\mu'+\sqrt{(1-\mu^2)(1-\mu'{}^2)}\cos(\phi-\phi').
\end{equation}
and combine this expression with~(\ref{angdist}) and~(\ref{hg}), to
obtain
\begin{equation}
	\Psi(\mu,\mu')
	=
	\frac{1-g^2}{\pi}
	\int_0^{\pi}\dfrac{\txd t}{\left(a\pm b\cos t\right)^{3/2}}
\end{equation}
with
\begin{gather}
	a(\mu,\mu') = 1+g^2-2g\mu\mu' 
	\\
	b(\mu,\mu') = 2|g|\sqrt{(1-\mu^2)(1-\mu'{}^2)}
\end{gather}
For all values of the asymmetry parameter, these functions obey the
relation $a>b\geq0$ such that we find (Gradshteyn \& Ryzhik, 1965,
Section~2.575)
\begin{equation}
	\Psi(\mu,\mu') 
	=
	\frac{2(1-g^2)}{\pi(a-b)\sqrt{a+b}}
	E\left(\sqrt{\frac{2b}{a+b}}\right),
\end{equation}
where $E(k)$ is the complete elliptic integral of the second
kind. Figure~{\ref{phasefun.ps}} shows the Henyey-Greenstein ARF at
the $V$ band for a few values of the initial direction $\mu'$.

\bsp

\begin{thebibliography}{}

\bibitem[2000]{AXBDK00} Alton P. B., Xilouris E. M., Bianchi S.,
Davies J., Kylafis N., 2000, A\&A, 356, 795

\bibitem[2001a]{BD01a} Baes M., Dejonghe H., 2001a, in {\em Galaxy Discs
and Disc Galaxies}, Funes J. G. and Corsini E. M. eds., ASP Conference
Series, in press

\bibitem[2001b]{BD01b} Baes M., Dejonghe H., 2001b, submitted to MNRAS

\bibitem[1997]{BQPS97} Berlind A. A., Quillen A. C., Pogge R. W.,
Sellgren J., 1997, AJ, 114, 107

\bibitem[1996]{BFG96} Bianchi S., Ferrara A., Giovanardi C., 1996,
ApJ, 465, 127

\bibitem[2000]{BFDA00} Bianchi S., Ferrara A., Davies J. I., Alton P. B.,
2000, MNRAS, 311, 601

\bibitem[1995]{BG95} Boselli A., Gavazzi G., 1994, A\&A, 283, 12

\bibitem[1992]{BBFA92} Bosma A., Byun Y., Freeman K. C., Athanassoula
E., 1992, ApJ, 400, L21

\bibitem[1988]{BMC88} Bruzual A. G., Magris C. G., Calvet N., 1988,
ApJ, 333, 673

\bibitem[1998]{BB98} Buat V., Burgarella D., 1998, A\&A, 334, 772

\bibitem[1991]{BHF91} Burnstein D., Haynes M., Faber S., 1991, Nature,
353, 515

\bibitem[1994]{BFK94} Byun Y. I., Freeman K. C., Kylafis N. D., 1994, ApJ,
432, 114

\bibitem[1994]{CKS94} Calzetti D., Kinney A. L., Storchi-Bergmann T.,
1994, ApJ, 429, 582

\bibitem[1959]{CE59} Cashwell E. D., Everett C. J., 1959, A Practical
Manual on the Monte Carlo Method for random Walk Problems, Pergamon,
New York

\bibitem[1960]{C60} Chandrasekhar S., 1960, Radiative Transport, Dover
Publications, New York

\bibitem[1996]{CBS96} Corradi R. L. M., Beckman J. E., Simonneau E.,
1996, \mbox{MNRAS}, 282, 1005

\bibitem[1995]{DB95} Davies J. I., Burnstein D., 1995, The Opacity of
Spiral Disks, Kluwer Academic Publishers, Dordrecht

\bibitem[1957]{D57} Davison B., 1957, Neutron Transport Theory,
Clarendon Press, Oxford

\bibitem[1996]{dJ96} de Jong R. S., 1996, A\&A, 313, 377

\bibitem[1995]{DBP95} Di Bartolomeo A., Barbaro G., Perinotto M.,
1995, MNRAS, 277, 1279

\bibitem[1989]{DDP89} Disney M., Davies J., Phillipps S., 1989, MNRAS,
239, 939

\bibitem[1996]{FBDG96} Ferrara A., Bianchi S., Dettmar R.-J.,
Giovanardi C., 1996, ApJS, 123, 437

\bibitem[FHG96]{FHG96} Firmani C., Hernandez X., Gallagher J., 1996,
A\&A, 308, 403

\bibitem[1980]{FRR80} Flannery B. P., Roberge W., Rybicki G. B., 1980,
ApJ, 236, 598

\bibitem[1970]{F70} Freeman K. C., 1970, ApJ, 160, 811

\bibitem[1994]{GHSWDF94} Giovanelli R., Haynes M. P., Salzer J. J.,
Wegner G., Da Costa L. N., Freudling W., 1994, AJ, 107, 2036

\bibitem[1965]{GR65} Gradshteyn I. S., Ryzhik I. M., 1965, Table of
Integrals, Series and Products, Academic Press Inc., New York

\bibitem[1987]{GR87} Guiderdoni B., Rocca-Volmerange B., 1987, A\&A,
186, 1

\bibitem[1993]{JP93} James P. A., Puxley P. J., 1993, Nature, 363, 240

\bibitem[1994]{JKBPH94} Jansen R. A., Knapen J. H., Beckman J. E.,
Peletier R. F., Hes R., 1994, MNRAS, 270, 373

\bibitem[1937]{H37} Henyey L. G., 1937, ApJ, 85, 107

\bibitem[1941]{HG41} Henyey L. G., Greenstein J. L., 1941, ApJ, 93, 70

\bibitem[1998]{KTGW98} Kuchinski L. E., Terndrup D. M., Gordon K. D.,
Witt A. N., 1998, AJ, 115, 1438

\bibitem[1987]{KB87} Kylafis N. D., Bahcall J. N., 1987, ApJ, 317, 637

\bibitem[1994]{MP94} Maccioni A., Perinotto M., 1994, A\&A, 284, 241

\bibitem[2000]{MW00} Matthews L. D., Wood K., 2000, astro-ph/0010033

\bibitem[1970]{M70} Mattila K., 1970, A\&A, 9, 53

\bibitem[1975]{MSM75} Milkey R. W., Shine R. A., Mihalas D., 1975,
ApJ, 217, 425

\bibitem[2000]{MKPX00} Misiriotis A., Kylafis N. D., Papamastorakis
J., Xilouris E. M., 2000, A\&A, 353, 117

\bibitem[1995]{OK95} Ohta K., Kodaira K., 1995, PASJ, 47, 17

\bibitem[1995]{PVMFKB95} Peletier R. F., Valentijn E. A., Moorwood
A. F. M., Freudling W., Knapen J. H., Beckman J. E., 1995, A\&A, 300,
L1

\bibitem[1998]{PR98} Pizagno J., Rix H.-W., 1998, AJ, 116, 2191
 
\bibitem[2000]{PDLS00} Pohlen M., Dettmar R.-J., L\"utticke R.,
Schwarzkopf U., 2000, A\&AS, 144, 405

\bibitem[1989]{P89} Press W. H., Flannery B. P., Teukolsky S. A.,
Vetterling W. T., 1989, Numerical Recipes, Cambridge University Press,
Cambridge

\bibitem[1997]{RS97} R\"onnback J., Shaver P. A., 1997, A\&A, 322, 38

\bibitem[1983]{R83} Roberge W. G., 1983, ApJ, 275, 292

\bibitem[1990]{SY90} Saio H., Yoshii Y., 1990, ApJ, 363, 40

\bibitem[2000]{SD00} Schwarzkopf U., Dettmar R.-J., 2000, A\&A, 361, 451

\bibitem[1990]{V90} Valentijn E. A., 1990, Nature, 346, 153

\bibitem[1994]{V94} Valentijn E. A., 1994, MNRAS, 266, 614

\bibitem[1997]{VAK97} Vansevi\v{c}ius V., Arimoto N., Kodaira K.,
1997, ApJ, 474, 623

\bibitem[1970]{vdHdJ69} van de Hulst H. C., de Jong T., 1969, Physica,
41, 151

\bibitem[1981]{vdKS81} van der Kruit P. C., Searle L., 1981, A\&A, 95, 105 

\bibitem[1989]{WFH89} Wainscoat R. J., Freeman K. C., Hyland A. R.,
1989, ApJ, 337, 163

\bibitem[1992]{WK92} White R. E. III, Keel W. C., 1992, Nature, 346, 153

\bibitem[2000]{KWC00} White R. E. III, Keel W. C., Conselice C. J.,
2000, ApJ, 542, 761

\bibitem[1977]{W77} Witt A. N., 1977, ApJS, 35, 1

\bibitem[1996]{WG96} Witt A. N., Gordon K. D., 1996, ApJ, 463, 681

\bibitem[2000]{WG00} Witt A. N., Gordon K. D., 2000, ApJ, 528, 799

\bibitem[1992]{WTC92} Witt A. N., Thronson H. A. Jr., Capuano
J. M. Jr., 1992, ApJ, 393, 611

\bibitem[1997]{WJ97} Wood K., Jones T. J., 1997, AJ, 114, 1405

\bibitem[1997]{XKPPH97} Xilouris E. M., Kylafis N. D., Papamastorakis
J., Paleologou E. V., Haerendel G., 1997, A\&A, 325, 135

\bibitem[1998]{XADKPT} Xilouris E. M., Alton P., Davies J., Kylafis
N. D., Papamastorakis J., Trewhella M., 1998, A\&A, 331, 894

\bibitem[1999]{XBKPP99} Xilouris E. M., Byun Y., Kylafis N. D.,
Paleologou E. V., Papamastorakis J., 1999, A\&A, 344, 868

\bibitem[1996]{XH96} Xu C., Helou G., 1996, ApJ, 456, 163

\bibitem[1984]{YMW84} Yusef-Zadeh F., Morris M., White R. L., 1984,
ApJ, 278, 186



\end{thebibliography}
\end{document}